\documentclass[twocolumn,twoside,slac_two]{revtex4}
\usepackage{graphicx}
\usepackage{fancyhdr}
\usepackage{url}
\newcommand{\fermi}{{\it Fermi}}

\begin{document}

\title{Gamma-ray light curve and VLBI polarization connection in Mrk 421}

%

\author{M. Giroletti$^a$, F. D'Ammando$^{a,b}$, M. Orienti$^a$, D. Paneque$^c$ on behalf of the \fermi-LAT collaboration\\
R. Lico$^{a,b}$, G. Giovannini$^{a,b}$, J.\,L. G\'omez$^d$, S. Jorstad$^e$, A. Marscher$^e$}
\affiliation{$^a$INAF Istituto di Radioastronomia, Bologna, 40129, Italy\\ 
$^b$University of Bologna, Department of Physics and Astronomy, Bologna, 40127, Italy\\
$^c$Max-Planck-Institut f\"ur Physik, F\"ohringer Ring 6, 80805 M\"unchen, Germany\\
$^d$Instituto de Astrof\'{\i}sica de Andalucia, IAA-CSIC, Apdo. 3004, 18080 Granada, Spain\\
$^e$Institute for Astrophysical Research, Boston University, Boston, MA 02215, USA}
%
%
%
%

\begin{abstract}
We present \fermi-LAT and multi-frequency, multi-epoch VLBA data for the TeV blazar Mrk\,421. We collected the data during a long and intensive multi-frequency campaign in 2011. We study the gamma-ray light curve, the photon index evolution and their connection to the radio data on sub-parsec scales, including total intensity, polarized flux density, polarization angle, spectral index, and rotation measure both in the core and the jet region. The VLBA data were obtained at 15 and 24 GHz for 12 epochs and at 43 GHz for 23 epochs, thus providing the best temporal and spatial coverage in the radio band ever achieved for a TeV blazar. We provide significant constraints on the jet Doppler factor, the presence of proper motion, the magnetic field configuration, and an intriguing connection between variability in the radio data and the gamma-ray light curve: the total intensity and polarized core emission reach a peak simultaneously to the main gamma-ray peak, followed by a rotation of the polarization angle at low frequency. Opacity-related, long wavelength polarization swings are also detected later in the year, possibly related to secondary peaks in the gamma-ray light curve, setting constraints on the physics of the gamma-ray zone.

\end{abstract}

\maketitle

\thispagestyle{fancy}


\section{INTRODUCTION}

Multi-wavelength variability studies provide an extraordinary opportunity to break degeneracies between the various blazar emission models \cite{Tavecchio1998,Maraschi1999,Krawczynski2001}, which predict flux variations (at a given energy band) with particles of different energies, cooling times, and cross sections for different processes \cite{Krawczynski2002}.
However, the complexity in resolving the underlying processes occurring in blazars can only be achieved through a well-sampled dedicated monitoring from radio to gamma rays lasting several years. Indeed, some of the latest and most interesting results on blazars come precisely from variability studies from well-sampled coordinated years-long multi-instrument observations \cite{Abdo2010,Marscher2010,Jorstad2010}. Unluckily, these multi-instrument observations were performed on flat spectrum radio quasars, but not on TeV BL Lac objects \citep[with the exception of BL Lacertae,][]{Marscher2008}, perhaps owing to the fact that these sources were challenging to observe in radio and gamma rays. The advent of \fermi-LAT and the current generation of Cherenkov telescopes (H.E.S.S., MAGIC and VERITAS) permits the accurate and systematic (regardless of the activity level) monitoring of the high energy bump, where a large fraction of the blazar emission is produced. In addition, recent works have shown that TeV blazars can be successfully imaged with the VLBA at up to 43 GHz, revealing the core and the inner jet structure with great accuracy \cite{Piner2010}. Therefore, the currently available instrumentation allows to study these sources with an unprecedented level of detail, which has the potential to shed light on the understanding of these complex objects.

In order to address the challenge involved in breaking the degeneracy between models, we have organized the most ambitious multifrequency monitoring to date of the famous TeV BL Lac object Mrk\,421, covering sub-mm (SMA), optical/IR (GASP), UV/X-ray ({\it Swift}, RXTE, MAXI), and gamma rays (\fermi-LAT, MAGIC, VERITAS). We summarize here the main results of the multi-frequency, multi-epoch, full polarization VLBA observations and of the gamma-ray \fermi-LAT monitoring through 2011 \cite{Lico2012,Blasi2013,Lico2014}. 

Throughout this paper, we adopt  $H_0=70$ km sec$^{-1}$ Mpc$^{-1}$, $\Omega_M=0.25$ and $\Omega_\Lambda=0.75$, in a flat Universe, so that at the redshift of Mrk\,421 ($z=0.031$), $1\, \mathrm{mas}=0.59\, \mathrm{pc}$. Spectral and photon indexes $\alpha$ and $\Gamma$ are defined such that the radio flux density and the gamma-ray photon flux are proportional to $\nu^{-\alpha}$ and $E^{-\Gamma}$, respectively; angles are measured from north through east. 

\section{OBSERVATIONS}

\begin{figure}
\centering
\includegraphics[width=0.8\columnwidth]{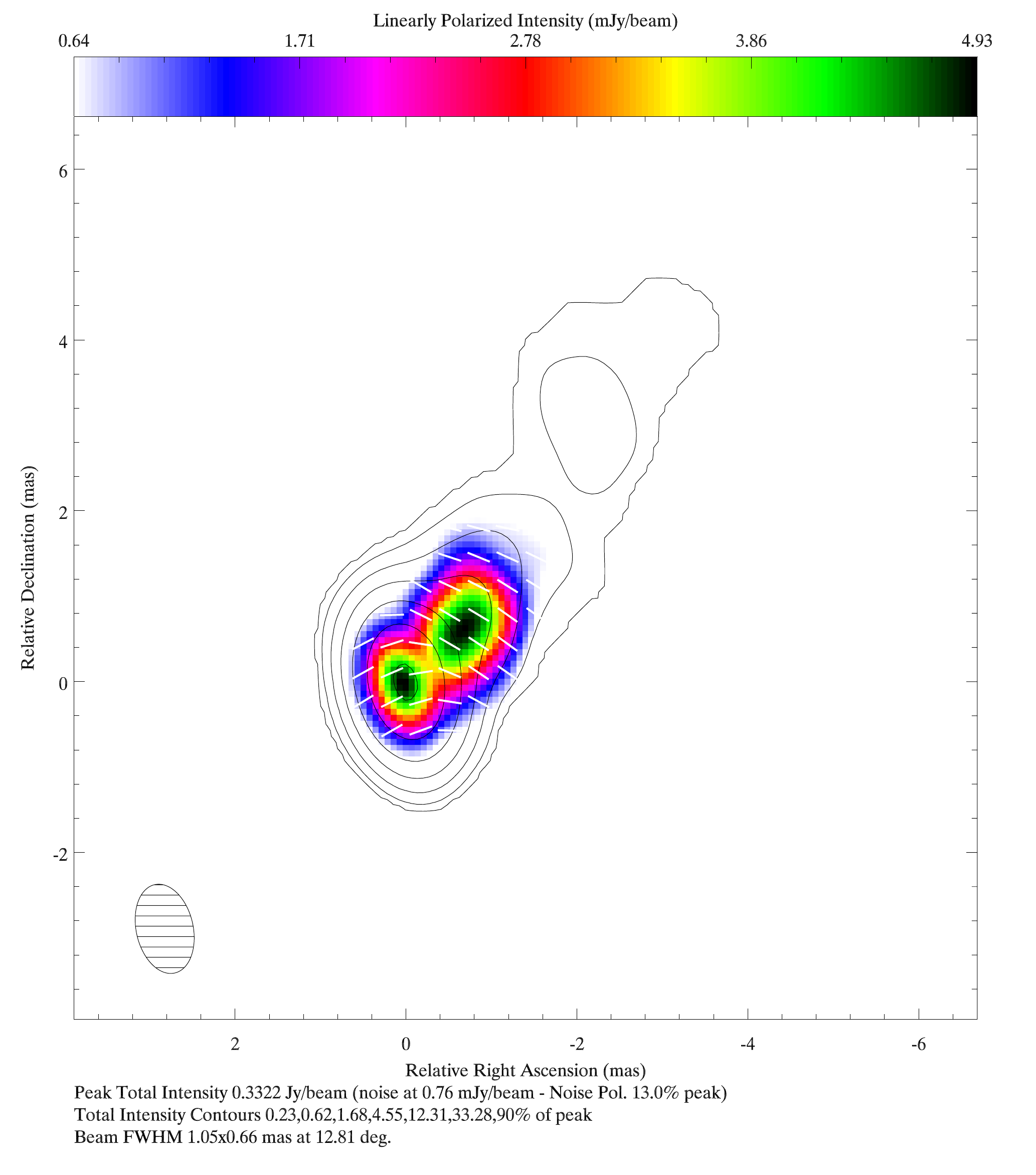}
\includegraphics[width=0.8\columnwidth]{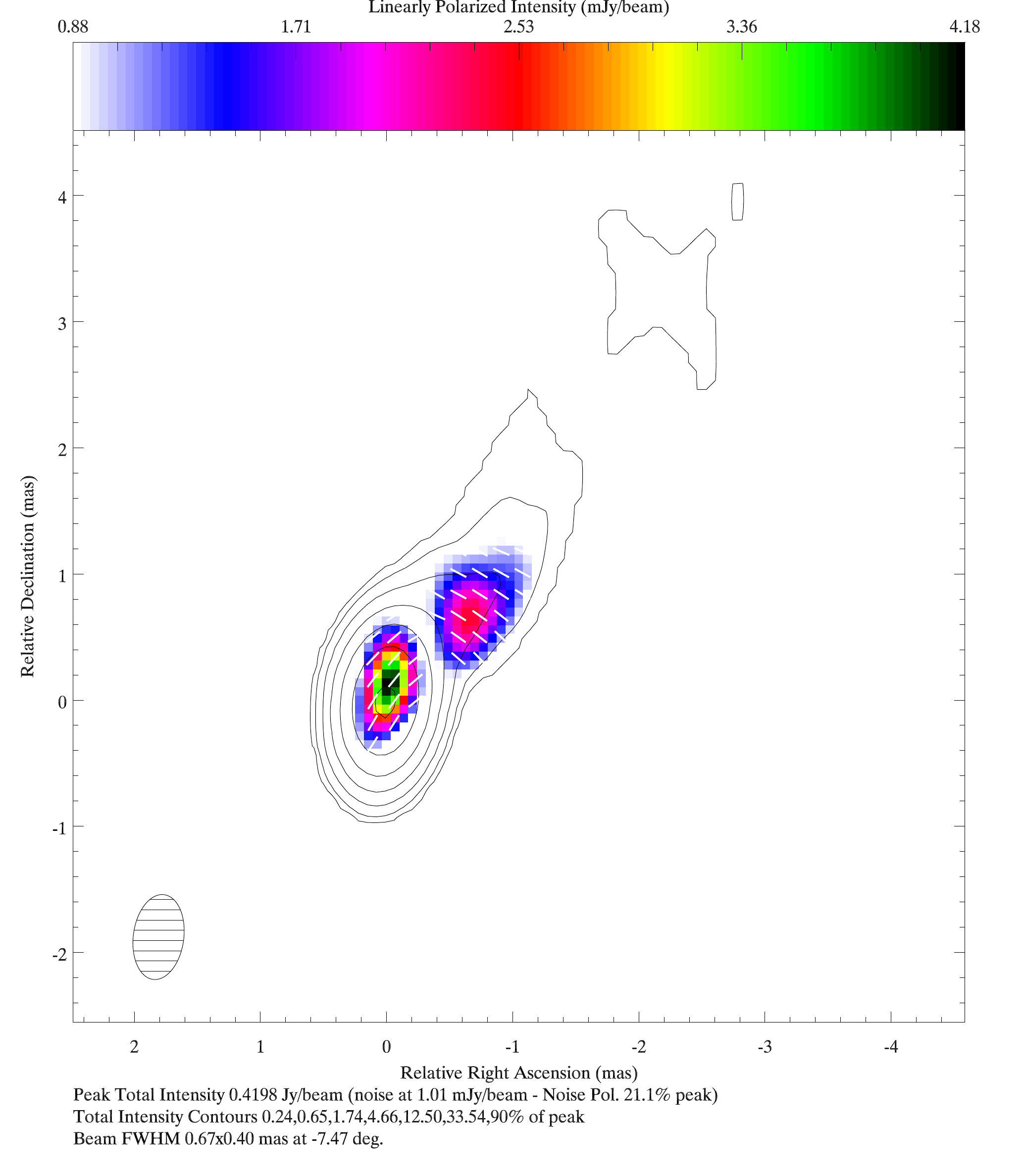}
\includegraphics[width=0.8\columnwidth]{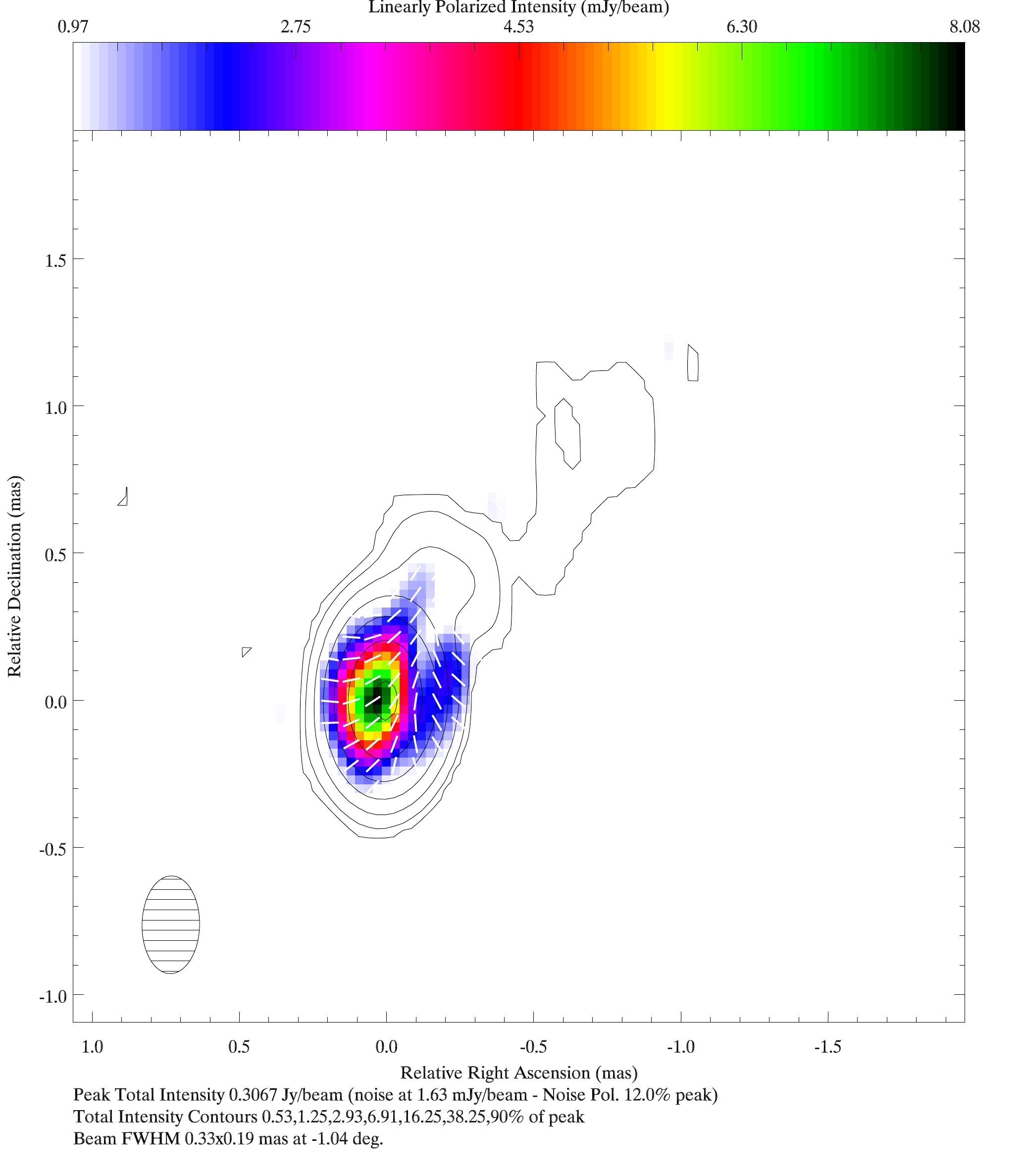}
\caption{VLBA images of Mrk\,421 at 15 GHz (top, 2011 January), 24 GHz (middle, 2011 February), and 43 GHz (bottom, 2011 April). }\label{f.images}
\end{figure}

\begin{figure}
\centering
\includegraphics[width=\columnwidth]{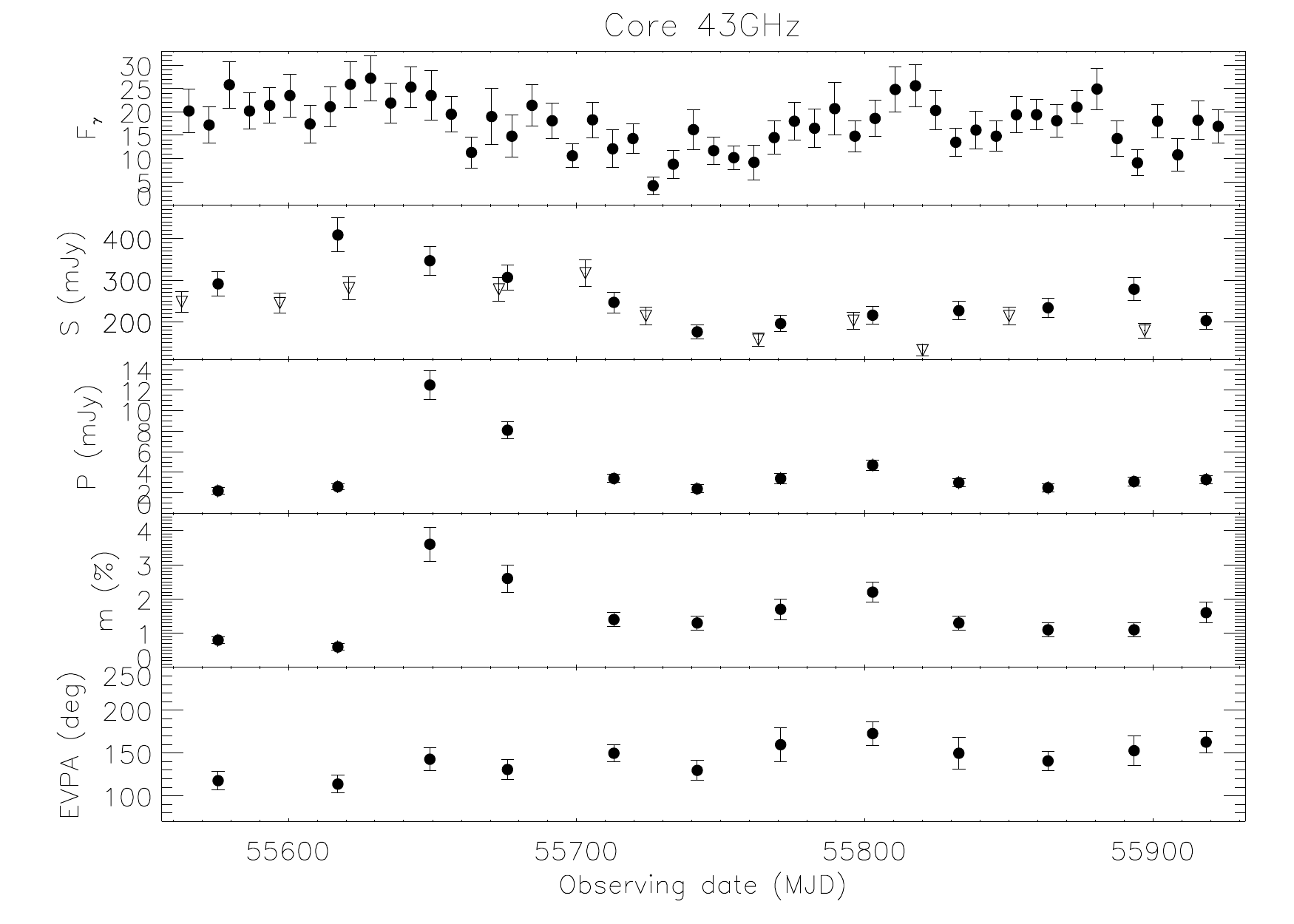}
\caption{Mrk\,421 43 GHz core parameters as a function of time. From top to bottom panel: gamma-ray light curve (0.1 to 100 GeV, in $10^{-8}$ ph cm$^{-2}$ s$^{-1}$ units, shown for comparison), total intensity flux density, polarized flux density, fractional polarization, EVPA. In the total flux density panel (second panel from top), filled circles indicate data from the main set of observations, empty triangles indicate additional data from the VLBA-BU-BLAZAR monitoring project.}\label{f.core}
\end{figure}

\subsection{Radio observations}
We observed Mrk\,421 with the VLBA for 12 times (once per month) throughout 2011 at 15, 24, and 43 GHz; at the latter frequency, we expanded our dataset with 11 more observations obtained through the VLBA-BU-BLAZAR monitoring program. We carried out a full calibration and analysis describing the evolution with time of total intensity and polarized flux density, and of their combinations such as spectral index and rotation measure. Core and inner jet are spatially resolved and separately analyzed. Spatial resolution is as good as 0.2 pc at 43 GHz and sensitivity just below 1 mJy/beam.

\subsection{Gamma-ray observations}

We analyzed gamma-ray data from the Large Area Telescope on board \fermi, which continuously scans the whole sky in the energy range 100 MeV $< E < $300 GeV. We analyzed the data with the ScienceTools software package version v9r32p5, using instrument response functions P7REP\_SOURCE\_V15 and following the standard procedures\footnote{\url{http://fermi.gsfc.nasa.gov/ssc/data/analysis/documentation/Cicerone/}}. Mrk\,421 is bright enough to be significantly detected in every weekly bin, and we obtained photon flux and photon index for every bin. 

\section{RESULTS}

We present sample images at the three frequencies in total intensity and polarization in Figure \ref{f.images}. In the following subsections, we describe the main results about the radio emission from the core (Sect.~\ref{s.core}) and jet (Sect.~\ref{s.jet}), and the unresolved gamma-ray source (Sect.~\ref{s.gamma}).

\begin{figure}
\centering
\includegraphics[width=\columnwidth]{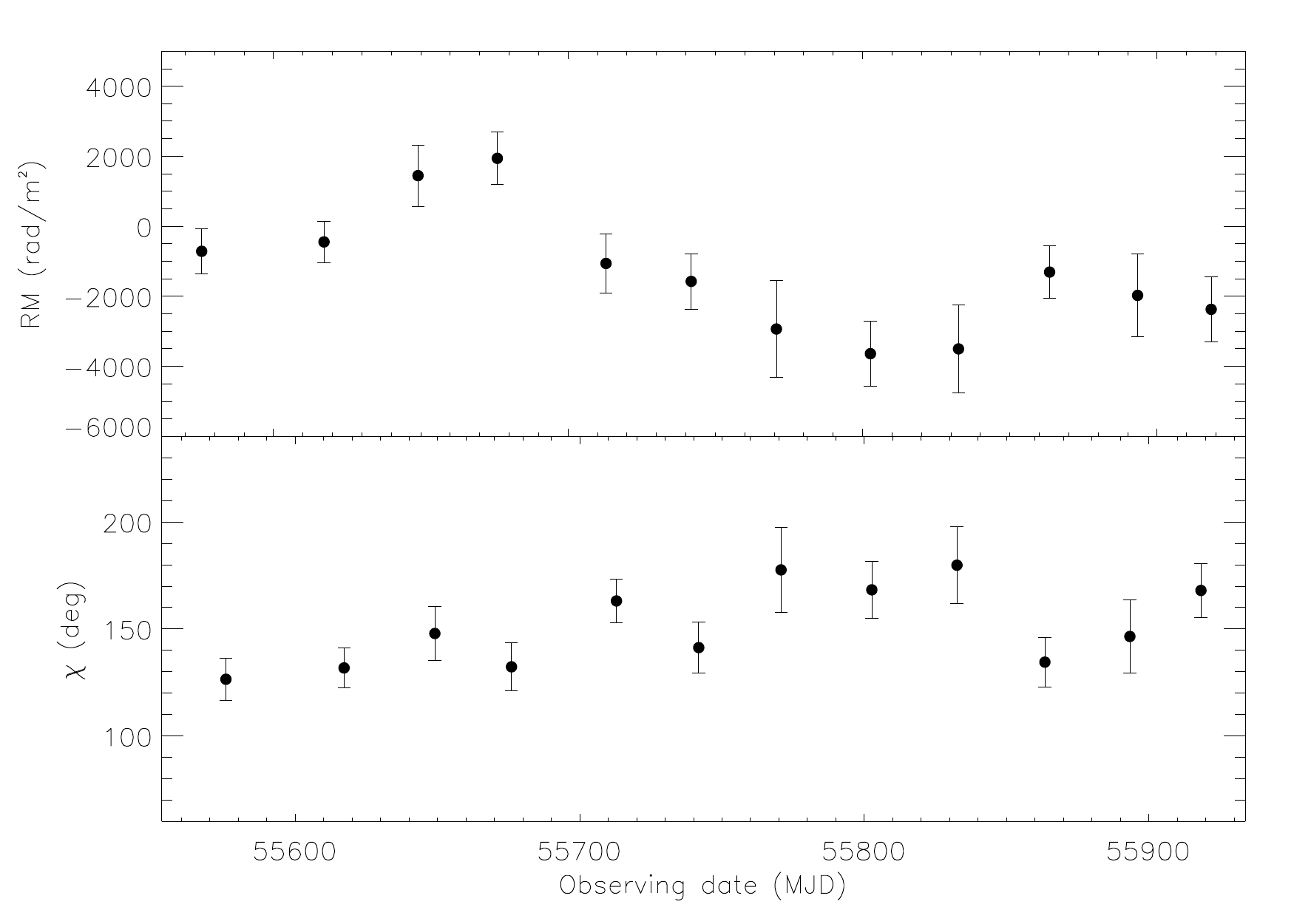}
\caption{Mrk\,421 core rotation measure (top panel) and intrinsic EVPA (bottom panel) as a function of time throughout 2011.} \label{f.rm}
\end{figure}

\subsection{Radio core\label{s.core}}

The core is the most prominent feature in the radio images. It is bright and compact at all epochs and frequencies and it shows variability in total intensity and polarization. We show its light curve and the time evolution of its polarization properties in Figure \ref{f.core}, along with the gamma-ray light curve that is described later on (Sect.~\ref{s.gamma}). 

The radio core light curve showed a broad peak around 2011 February-March at all frequencies. This peak is particularly prominent at 43 GHz, thanks to the improved sampling offered by the additional 11 observations from the Boston University blazar monitoring project. In particular, the core flux density reached its peak of $S_{43}=415$ mJy  on 2011 March 1 at 43 GHz. The core spectrum is generally flat, with average spectral indexes $\alpha_{15}^{24}=0.16$ and $\alpha_{24}^{43}=0.43$. The spectral index shows a flatter when brighter behaviour.

The core polarization properties are also variable:\ the polarization fraction varies as a function of both  frequency (being $\sim 1\%$ at 15 and 24 GHz and $\sim 2\%$ at 43 GHz) and time, with a ($3.6\pm0.5)\%$ peak on 2011 March 29. The electric vector position angle (EVPA) $\chi$ has some fluctuations, which are more prominent at low than at higher frequency. At 43 GHz, it oscillates slightly with a mean value of $144^\circ$ and a standard deviation of $17^\circ$, being overall well aligned to the jet axis. At 15 GHz, we detect two prominent $90^\circ$ flips in July and September, most likely due to opacity; even after correcting these values, there is larger variability with a mean of $119^\circ$ and a standard deviation of $29^\circ$.

Epoch by epoch, we performed a $\chi$ vs $\lambda^2$ fit to determine the rotation measure RM and the intrinsic EVPA $\chi_0$ in the source, under the assumption that the frequency dependence of the observed EVPA's is due to Faraday rotation effect. The evolution of RM and $\chi_0$ is shown in Figure \ref{f.rm}. The RM values oscillate between $-3000$ and $2000\ \mathrm{rad}\ \mathrm{m}^{-2}$. 


\subsection{Radio jet\label{s.jet}}

\begin{figure}
\centering
\includegraphics[width=\columnwidth]{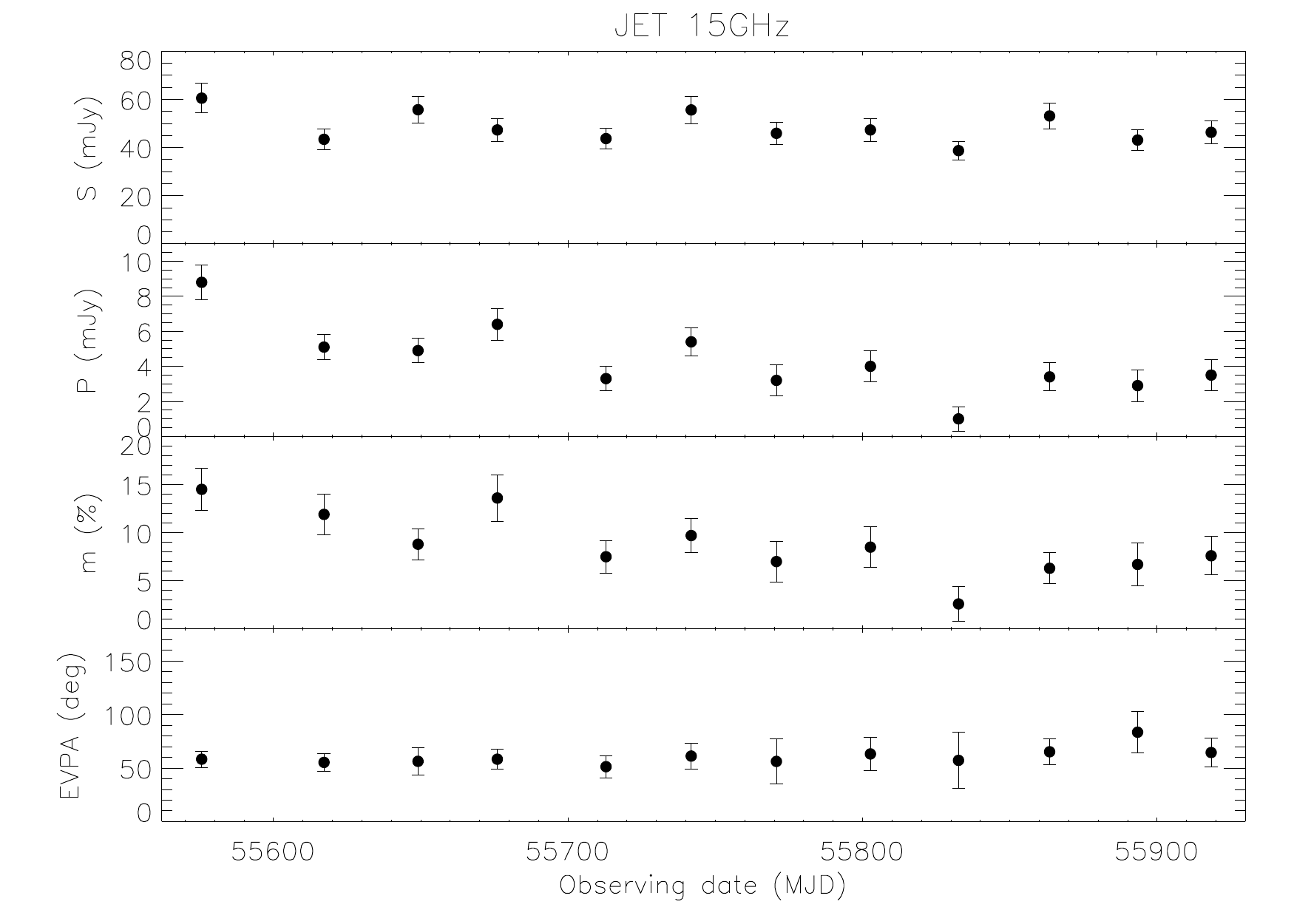}
\caption{Mrk\,421 15 GHz jet parameters as a function of time. From top to bottom panel: total intensity flux density, polarized flux density, fractional polarization, EVPA.} \label{f.jet}
\end{figure}

The VLBA data clearly detect a one-sided jet at all three frequencies. The jet extends for about 5 mas (3 pc) in PA $-35^\circ$ at 15 GHz and somewhat less as frequency increases. In Figure \ref{f.jet}, we show the total intensity and polarized flux density, the fractional polarization, and the EVPA of the jet emission at 15 GHz, from top to bottom.

We do not detect any prominent variation in neither the jet's structure nor its total intensity flux density. In particular, we did not detect any significant superluminal motion of components \cite{Lico2012,Blasi2013}; as a matter of fact, we do not detect well defined, compact jet components at all. At a reference distance of 1.5 mas from the core, the jet-counterjet brightness ratio is $R>30$. The spectrum is steeper than in the core.

In polarization, we detect emission from the jet at 15 and 24 GHz, with a fractional polarization value that is much larger than in the core ($\sim 15\%$). However, given the steep spectrum of the jet and its low surface brightness, at 43 GHz the polarized emission is below the noise level. The EVPA in the jet is remarkably stable, with a mean of $61^\circ$ and a standard deviation of just $9^\circ$. Given the jet axis PA of $-35^\circ$, this means that the EVPA is nearly orthogonal to the direction of the flow, and the magnetic field well aligned with it.


\subsection{Gamma-ray source\label{s.gamma}}

\begin{figure}
\centering
\includegraphics[width=\columnwidth]{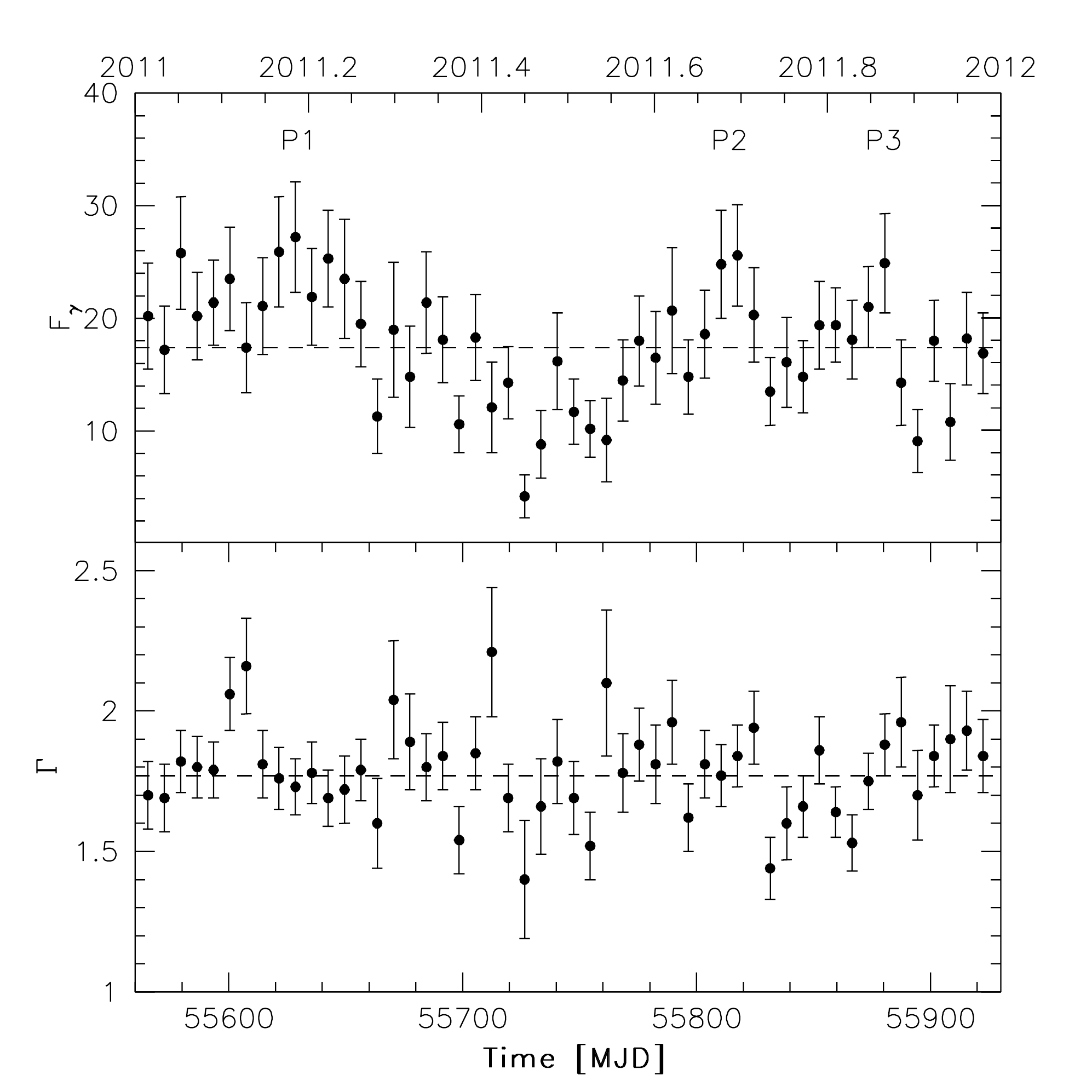}
\caption{Mrk\,421 gamma-ray parameters as a function of time. Top: photon flux; bottom: photon index. The dashed lines indicate the 1-yr mean value.} \label{f.gamma}
\end{figure}

Mrk\,421 is point like at the \fermi\ angular resolution. This is not only true for the angular scales probed by the VLBA observations presented here but also when we consider the total extent of the source as mapped by compact interferometers best suited to map a possible large scale diffuse emission \cite{Giroletti2006}. However, the variability time scales observed for the high energy emission and the broadband spectral energy distribution modelling \cite{Abdo2011} clearly suggest that most of the gamma-ray photons are produced in a compact region of size smaller than the radio core itself.

During 2011, the mean photon flux and the photon index are $F =(17.4\pm0.5)\times10^{-8}\ \mathrm{ph}\ \mathrm{cm}^{-2}\ \mathrm{s}^{-1}$ and $\Gamma=1.77\pm0.02$, respectively. The light curve on weekly time bins shown in Figure \ref{f.gamma} (top panel) reveals variability, with three peaks in 2011 March, September, and November. The brightest week is that between 2011 March 5 and 11, (MJD 55\,625-55\,631), with a gamma-ray flux as large as $(38\pm11)\times10^{-8}\ \mathrm{ph}\ \mathrm{cm}^{-2}\ \mathrm{s}^{-1}$. The photon index on the other hand is quite stable and we do not find any evidence for spectral variability (see Figure \ref{f.gamma}, bottom panel). 

\section{DISCUSSION}

The results from the radio observations suggest that the jet is not strongly Doppler boosted already on parsec scale. This stems from the low brightness temperature estimated both from the variability time scales ($T_\mathrm{B,\,var} \sim 2.1 \times 10^{10}$ K) and the core size and flux density ($T_\mathrm{B} \sim 10^{11}$ K) \cite{Lico2012}. On the contrary, TeV variability on time scales of about 30 minutes \cite{Gaidos1996}, and both hadronic (synchrotron proton blazar) or leptonic (one-zone synchrotron self-Compton scenario) fits to the broadband spectral energy distribution require Doppler factors $\delta$ in excess of $\sim 10$, and possibly as large as $\delta = 50$ \cite{Abdo2011}. This so-called ``Doppler factor crisis'' characterizes also other TeV blazars; the most natural way to solve this crisis is to localize the radio and the gamma-ray emission zones in different regions; typically, a velocity structure of the jet is assumed, either along or across the jet axis.

\begin{figure}
\centering
\includegraphics[width=\columnwidth]{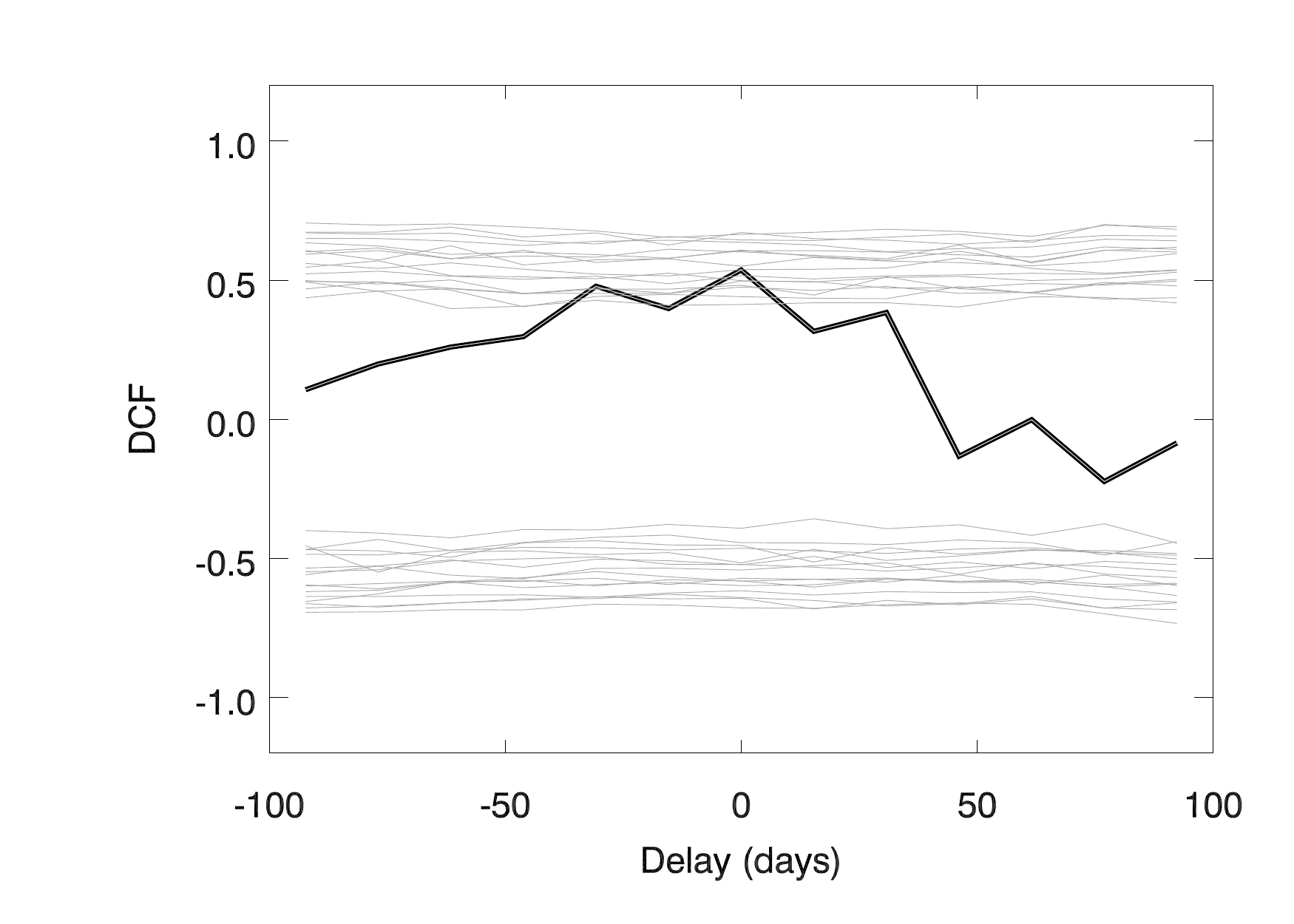}
\caption{DCF results for Mrk\,421, considering gamma-ray \fermi\ data and  43 GHz VLBA data. Black line: observed data; gray curves: 99.7\% confidence level threshold obtained from the combination of different PSD slopes.} \label{f.dcf}
\end{figure}

In our case, we have the possibility to compare the radio and gamma-ray light curves and to carry out a discrete cross-correlation function (DCF) analysis for the two datasets. In particular, we carried out a DCF analysis over a range of radio-gamma ray delays between $-100$ and +100 days, with a bin of 15 days. We find a peak for the correlation (0.54) is obtained for $\Delta t=0$ day delay. In order  to assess the significance of this peak, we generated 3000 fake light curves with the same mean and standard deviation as the observed light curves but with variable power spectral densities (PSD). We show in Figure \ref{f.dcf} the results of our DCF analysis on the real data compared to those obtained for various combinations of the simulated ones. In particular, the peak is significant at the $>99.7\%$ confidence level for the combinations of light curves with short time scale gamma-ray variability ($\beta_{\gamma\mathrm{-ray}}<1.5$) and longer time scale radio variability ($1<\beta_\mathrm{r}<2.5$).

As for the magnetic field configuration in the radio core and jet, we found somewhat discrepant results. The core B-field perpendicular to the jet axis is typical of shocked regions; in the jet, the magnetic field is aligned with the jet axis; possible interpretations of this result include stretching of an initially transverse magnetic field by a layered velocity structure and an helical field with a small pitch angle. Finally, we note that the variability features in the polarized flux light curve indicate also a connection between magnetic field and gamma-ray emission.


\bigskip 
\begin{acknowledgments}
The \fermi-LAT Collaboration acknowledges support for LAT development, operation and data analysis from NASA and DOE (United States), CEA/Irfu and IN2P3/CNRS (France), ASI and INFN (Italy), MEXT, KEK, and JAXA (Japan), and the K.A. Wallenberg Foundation, the Swedish Research Council and the National Space Board (Sweden). Science analysis support in the operations phase from INAF (Italy) and CNES (France) is also gratefully acknowledged.
 
We acknowledge financial contribution from grant PRIN-INAF-2011. Part of this work was done with the contribution of the Italian Ministry of Foreign Affair. Part of this work was supported by the Marco Polo program of the University of Bologna and the COST Action MP0905 "Black Holes in a Violent Universe". The research at the Instituto de Astrofisica de Andalucia was supported in part by the Spanish Ministry of Economy and Competitiveness grant AYA2010-14844 and by the Regional Government of Andaluc\'ia (Spain) grant P09-FQM-4784.  The research at Boston University was supported in part by NASA through Fermi grants NNX08AV65G, NNX08AV61G, NNX09AT99G, NNX09AU10G, and NNX11AQ03G, and by US National Science Foundation grant AST-0907893.


\end{acknowledgments}

\bigskip 

\end{document}